\documentclass[prd,eqsecnum]{revtex4}
\usepackage{graphicx}
\usepackage{dcolumn}
\usepackage{bm}
\usepackage{axodraw}

\newcommand{\be}{\begin{equation}}
\newcommand{\ee}{\end{equation}}
\newcommand{\bse}{\begin{subequations}}
\newcommand{\ese}{\end{subequations}}
\newcommand{\bary}{\begin{eqnarray}}
\newcommand{\eary}{\end{eqnarray}}

\newcommand{\gmu}{\gamma^{\mu}}
\newcommand{\gnu}{\gamma^{\nu}}
\newcommand{\Ee}{\langle E_e\rangle}
\newcommand{\Eebar}{\langle E_{\bar e}\rangle}
\newcommand{\Enu}{\langle E_{\nu}\rangle}
\newcommand{\iEe}{\langle \frac{1}{E_e}\rangle}
\newcommand{\iEebar}{\langle \frac{1}{E_{\bar e}}\rangle}
\newcommand{\iEf}{\langle \frac{1}{E_f}\rangle}

\begin{document}

\title{Higher order correction to the neutrino self-energy in a 
medium and its astrophysical applications}

\author{ALBERTO BRAVO GARC\'IA and SARIRA SAHU}
\affiliation{
Instituto de Ciencias Nucleares, Universidad Nacional Aut\'onoma de
M\'exico,
Circuito Exterior, C. U., A. Postal 70-543, 04510,
Mexico D.F., Mexico}

\begin{abstract}
We have calculated the $1/M^4$ ($M$ the vector boson mass) 
order correction to the neutrino self-energy
in a medium. The possible application of this higher order contribution to
the neutrino effective potential is considered in the context of the Early 
Universe
hot plasma and of the cosmological Gamma Ray Burst fireball. We found that, 
depending on the medium parameters and on the neutrino properties (mixing angle
and mass square difference)
the resonant oscillation of active to active neutrinos is possible.

\end{abstract}
\maketitle

\section{Introduction}
The particle propagation in the presence of a finite temperature and density
medium has been studied extensively because of its importance in various 
astrophysical and cosmological scenarios. The vacuum dispersion relation 
is no longer respected in the medium and the properties of the particle
are modified. This has been studied in great detail for photon and charged 
particles$^1$. The neutrino propagation in a medium was first 
studied by Wolfenstein, who determined the refractive index of $\nu_e$ from the
forward scattering amplitude of $\nu_e$-electron scattering$^2$. 
Using the finite 
temperature field theory method, the neutrino propagation has been studied
in various environments$^{3-6}$.

In the medium, the test neutrino or anti-neutrino will interact with the
background particles through both charged current (CC) and neutral current (NC)
 interactions depending on the nature of the background. For example, the 
electron neutrino will interact both by CC and NC interaction with the 
background of normal matter, whereas the muon and tau neutrinos will interact
only through NC interaction in the normal matter, because of the absence
of muons and taus in the medium. However, exotic environments like, early 
Universe hot plasma and core of the supernovae have a substantial amount of 
neutrinos, trapped within it, which can contribute to the potential due to
$\nu-\nu$ scattering. The leading order contribution to 
the neutrino potential is proportional to the difference of the number 
densities of particles. So in an almost CP-symmetric medium 
(e. g. early Universe)
the leading order ($1/M^2$) contribution to the potential is
very small and one has to go beyond the leading order 
(to order $1/M^4$)$^{3,4}$. 

The early Universe must possess a tiny excess of matter
over antimatter i.e. $N_B/N_{\gamma}\simeq 5\times 10^{-10}$ to avoid 
total annihilation of matter-antimatter and charge
neutrality implies such an excess of electrons over 
positrons, but total lepton asymmetry may not be comparable to baryon
asymmetry$^7$. In fact larger lepton asymmetry is not 
being excluded by any theoretical reasons.
As it is well known up to present, neutrino oscillation in the early 
Universe can have dramatic consequences on BBN. The active-sterile neutrino
oscillation in the early Universe can create a large asymmetry between 
$\nu_e$ and ${\bar\nu}_e$ which could have a direct effect on the nuclear 
reactions and finally the abundances in the light 
elements$^{8-12}$. Similar situation may occur, in the
cosmological Gamma Ray Burst (GRB) fireball, where radiation and electron 
positron plasma is formed in a compact region with extremely high optical
depth, so that, photons can not escape from it and they split into 
electron positron
pairs. This fireball may be contaminated by baryons both from the 
central engine and the surrounding medium$^{13-18}$. 
Neutrinos from the progenitor as well as from the inverse beta decay
process within the fireball will propagate in the fireball. Depending on the 
nature of the fireball (e. g. temperature), there may be resonant oscillation 
of active neutrinos within the fireball$^{19}$.

In this paper we calculate the neutrino effective potential up to
order $1/M^4$ and consider its
effect on neutrino oscillation in different environments like early Universe
hot plasma$^{3,4}$  and the GRB fireball. We found that, 
depending on the parameters
of the fireball, there can be resonant oscillation from active to active
neutrino. On the contrary, the anti-neutrino conversion is very much 
suppressed. 
 
The paper is organized as follows: In sec.2 we give a detailed calculation
of the neutrino self-energy, from different diagrams that contribute to it.
Sec.3 is devoted to the calculation of neutrino effective potential
in different backgrounds. The 
oscillation of neutrinos in the 
early Universe hot plasma and the GRB fireball environment are discussed
in sec. 4 and a brief conclusion is drawn in Sec. 5.

\section{Neutrino self energy}

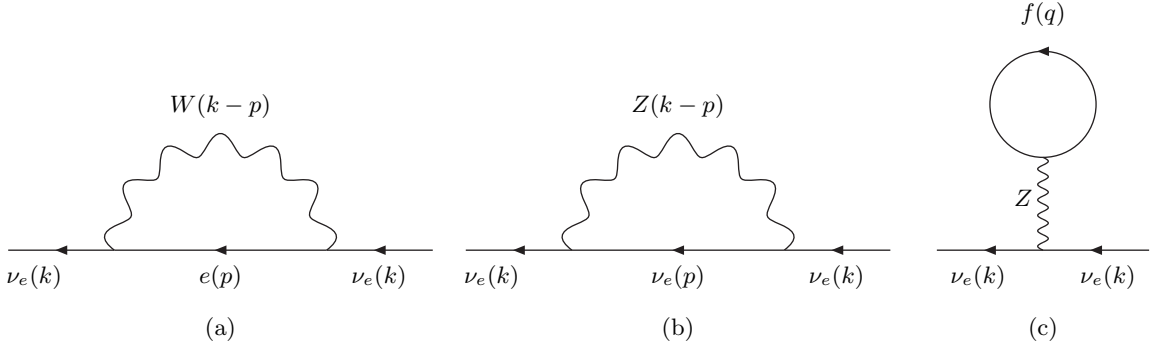
\begin{figure}
\begin{center}
%
%
\begin{picture}(170,130)(-85,-30)
\Text(0,-30)[c]{(a)}
\ArrowLine(80,0)(40,0)
\Text(60,-10)[c]{$\nu_e(k)$}
\ArrowLine(40,0)(-40,0)
\Text(0,-10)[c]{$e(p)$}
\ArrowLine(-40,0)(-80,0)
\Text(-60,-10)[cr]{$\nu_e(k)$}
\PhotonArc(0,0)(40,0,180){4}{6.5}
\Text(0,50)[cb]{$W(k-p)$}
\end{picture}
\begin{picture}(170,130)(-85,-30)
\Text(0,-30)[c]{(b)}
\ArrowLine(80,0)(40,0)
\Text(60,-10)[c]{$\nu_e(k)$}
\ArrowLine(40,0)(-40,0)
\Text(0,-10)[c]{$\nu_e(p)$}
\ArrowLine(-40,0)(-80,0)
\Text(-60,-10)[cr]{$\nu_e(k)$}
\PhotonArc(0,0)(40,0,180){4}{6.5}
\Text(0,50)[cb]{$Z(k-p)$}
\end{picture}
%
\begin{picture}(100,100)(-50,-30)
\Text(0,-30)[c]{(c)}
\ArrowLine(40,0)(0,0)
\Text(35,-10)[cr]{$\nu_e(k)$}
\ArrowLine(0,0)(-40,0)
\Text(-35,-10)[cl]{$\nu_e(k)$}
\Photon(0,0)(0,35){2}{6}
\Text(-4,20)[r]{$Z$}
\ArrowArc(0,55)(20,-90,270)
\Text(0,85)[b]{$f(q)$}
\end{picture}
\caption[]{One-loop diagrams for the neutrino self-energy 
in a medium.
\label{fig1}
}
\end{center}
\end{figure}

In the medium, the Dirac equation 
for the neutrino is given by $\left [ {\not p} - \Sigma(p)\right ]\psi =0 $,
where $p$ is the neutrino four momentum and $\Sigma(p)$ is its self-energy. 
The neutrino self-energy can be calculated from the three diagrams 
given in Fig.\ref{fig1}. The contribution due to the W-exchange 
Fig.\ref{fig1} (a)
is given by 
\be
-i\Sigma_W(k) =\frac{g^2_W}{2} R \int\frac{d^4p}{(2\pi)^4}
\left [\gamma_{\mu}S_e(p)\gamma_{\nu}D^{\mu\nu}(q)\right ] L,
\label{sfen}
\ee
where we have defined the four momentum $q=k-p$, $R$ and $L$ are the right 
and left projection 
operators defined as $R,L=\frac{1\pm \gamma^5}{2}$ and  $S_e(p)$ is 
the electron propagator given by
\be
S_e(p) = ({\not p}+m)\left (\frac{1}{(p^2-m^2)} + i\Gamma_e(p)\right ).
\label{fprop}
\ee
The vector boson propagator up to order $1/M^4$ is given by
\be
D^{W,Z}_{\mu\nu}(p) \simeq
\frac{1}{M_{W,Z}^2}
\left (g_{\mu\nu}+\frac{g_{\mu\nu}p^2}{M_{W,Z}^2}
-\frac{p_{\mu}p_{\nu}}{M_{W,Z}^2}\right ).
\label{bprop}
\ee
In the electron propagator
\be
\Gamma_e(p)=2\pi \delta (p^2-m^2)\eta(p.u),
\ee
and
\be
\eta(p.u) =\Theta (p.u) f(p.u) + \Theta (-p.u) f(-p.u),
\ee
where $u=(1,{\bf 0})$ is the four velocity of the medium, 
considered at rest,  and $f(p.u)$ is
the fermion distribution function.
The real part of Eq.(\ref{sfen}) is given by
\bary
Re\Sigma_W(k) &=&
-\frac{g^2_W}{2M^2_W}\int\frac{d^4p}{(2\pi)^3}\delta(p^2-m^2)\eta(p.u)
R\gamma_{\mu}({\not p}+m)\gamma_{\nu}L \times\nonumber\\
&\times& \left (g_{\mu\nu}+\frac{g_{\mu\nu}q^2}
{M^2_W}-\frac{q_{\mu}q_{\nu}}{M^2_W}\right ).
\label{rsw}
\eary
For simplicity let us define $Re\Sigma_W=R{\tilde\Sigma_W}L$
and by doing the $p_0$ integral, we get
\bary
{\tilde\Sigma_W}(k)
&=&\frac{g^2_W}{2M^2_W}\int \frac{d^3p}{(2\pi)^3 E}
\left[({\not p}-2m) f(p.u)-({\not p}+2m) {\bar f}(p.u)\right.\nonumber\\
&+&\left.
\frac{1}{M^2_W}
\left [
\left \{
q^2({\not p}-2m)
+\frac{q^{\mu}q^{\nu}}{2}
\gamma_{\mu}({\not p}+m)\gamma_{\nu}
\right \}f(p.u)\right.\right.\nonumber\\
&-&\left.\left. 
\left \{
q^2({\not p}+2m)
+\frac{q^{\mu}q^{\nu}}{2}
\gamma_{\mu}({\not p}-m)\gamma_{\nu}
\right \}{\bar f}(p.u)
\right] \right ].
\label{tsgm}
\eary

%
%
%
In general ${\tilde\Sigma}_W$ is a function of the external
four momentum $k^{\mu}=(\omega, {\bf k})$ and the four velocity $u^{\mu}$
of the heat bath. Then this can be expressed as 
\be
-{\tilde\Sigma}_W=a_W{\not k} + b_W{\not u},
\label{ls}
\ee
where $a_W$ and $b_W$ are Lorentz scalars and can be calculated from,
\be
T_k=-\frac{1}{4}Tr[{\not k}Re{\tilde\Sigma}_W]= a_W k^2 + b_W \omega,
\label{tk}
\ee
and
\be
T_u=-\frac{1}{4}Tr[{\not u}Re{\tilde\Sigma}_W]= a_W \omega + b_W.
\label{tu}
\ee
For a test neutrino of energy $E_{\nu}=k.u$ we obtain,
\be
a_W=\frac{E_{\nu}}{\bf k^2} T_u -\frac{T_k}{\bf k^2},
\label{aw1}
\ee
and
\be
b_W=T_u-a_W E_{\nu}.
\label{bw1}
\ee
For massless neutrinos $k^2=0$ which implies $E_{\nu}=|{\bf k}|$. 
Using Eq.(\ref{tsgm}) in Eqs.(\ref{tk}) and (\ref{tu}) we obtain
\bary
T_k&=&
-\frac{g^2_W}{2M^2_W}\int \frac{d^3p}{(2\pi)^3}
\left [
E_{\nu} \left (1+ \frac{3m^2}{2M^2_W}\right )(f(p.u)-{\bar f}(p.u))
\right.\nonumber\\
&&\left.-\frac{2 E^2_{\nu}}{3M^2_WE}
(4E_e^2-m^2)(f(p.u)+{\bar f}(p.u)) \right ], 
\label{tkk}
\eary
and 
\bary
T_u&=&-\frac{g^2_W}{M^2_W}\int \frac{p^2dp}{(2\pi)^2 E_e} 
\left [ 
E_e (f(p.u)-{\bar f}(p.u))+\frac{1}{M^2_W}
\left ( \frac{3m^2 E_e}{2} + E^2_{\nu} E_e\right ) 
(f(p.u)\right.
\nonumber\\
&&\left. -{\bar f}(p.u))-\frac{1}{M^2_W}
\left ( 2E_e^2 E_{\nu}+m^2 E_{\nu} \right ) (f(p.u)+{\bar f}(p.u))
\right ],
\label{tuu}
\eary
with $E_e$ the electron energy and $m$ the electron mass.
In Eqs.(\ref{tkk}) and (\ref{tuu}) we have three different types of terms
which are proportional to $f(p.u)$, $E_e f(p.u)$ and $f(p.u)/E_e$ and 
also proportional to corresponding anti-fermion distribution functions.
The number densities of fermions and
anti-fermions are defined as,
\be
N_f=g\int\frac{d^3p}{(2\pi)^3} {f}_f(p.u)~~~~\text{and}~~~~
{\bar N}_f=g\int\frac{d^3p}{(2\pi)^3} {\bar f}_f(p.u),
\label{fand}
\ee
respectively, where $g=2$ for $e, p~{\text and}~ n$ and one for neutrinos. 
For $E_e\gg\mu$, the fermion distribution
function can be written in terms of a series, as follows
\be
f(E_e)=\frac{1}{e^{\beta (E_e-\mu)}+1}=\sum_{l=1}^{\infty}
(-1)^{(l+1)} e^{\alpha_l} e^{-\beta E_e l},
\label{fdf}
\ee
where $\mu$ is the chemical potential and $\alpha_l=\beta\mu l$.
Using Eq.(\ref{fdf}) the number 
density can be written as
\be
N_f=gm^3 \sum_{l=0}^{\infty}(-1)^l e^{\alpha_l}
\int_1^{\infty} t dt \sqrt{t^2-1}e^{-a_lt}
=g\frac{m^3}{2\pi^2}\sum_{l=1}^{\infty}(-1)^{(l+1)} e^{\alpha_l}
\frac{{\cal K}_2(a_l)}{a_l},
\label{nd}
\ee
where $a_l=\beta ml$ and ${\cal K}_i,~~ {i=1,2,..}$ are the 
modified Bessel functions of 
integer order. Similarly, we have
\be
{\cal J}_2=\int \frac{d^3p}{(2\pi)^3} \frac{f(E_e)}{E_e}
=\frac{gm^3}{2\pi^2}\sum_{l=1}^{\infty}(-1)^{(l+1)} e^{\alpha_l}
\frac{{\cal K}_1(a_l)}{a_l},
\label{ndd}
\ee
and
\bary
{\cal J}_3&=&g\int \frac{d^3p}{(2\pi)^3} E_ef(E_e)
=\frac{gm^4}{2\pi^2}\sum_{l=1}^{\infty}(-1)^{(l+1)} e^{\alpha_l}
\times\nonumber\\
&\times& \left [
\frac{3}{a_l^2}{\cal K}_0(a_l) + \frac{{\cal K}_1(a_l)}{a_l} 
\left (1+\frac{6}{a_l^2}\right )\right ],
\label{nddd}
\eary
which gives
\be
N_f-{\bar N}_f = g\frac{m^3}{\pi^2}\sum_{l=1}^{\infty}(-1)^{(l+1)} 
\sinh(\alpha_l)\frac{{\cal K}_2(a_l)}{a_l},
\ee
\be
\Phi_2={\cal J}_2+{\bar {\cal J}}_2=
g\frac{m^2}{\pi^2}\sum_{l=1}^{\infty}(-1)^{(l+1)} 
\cosh(\alpha_l)\frac{{\cal K}_1(a_l)}{a_l},
\label{den}
\ee
and 
\be
\Phi_3={\cal J}_3+{\bar {\cal J}}_3=
g\frac{m^4}{\pi^2}\sum_{l=1}^{\infty}(-1)^{(l+1)} 
\cosh(\alpha_l)\left [
\frac{3}{a_l^2}{\cal K}_0(a_l) + \frac{{\cal K}_1(a_l)}{a_l} 
\left (1+\frac{6}{a_l^2}\right )\right ].
\label{denn}
\ee
Inserting these expressions in Eqs.(\ref{tkk}) and (\ref{tuu}), we obtain
\bary
T_k&=&-\frac{g^2_W}{2M^2_W}
\left [
E_{\nu}\left (1+ \frac{3m^2}{2M^2_W}\right )(N_e-{\bar N}_e)
\right.\nonumber\\
&&\left.
-\frac{8 E^2_{\nu}}{3M^2_W}\Phi_3
+\frac{2 E^2_{\nu} m^2}{3M^2_W}\Phi_2
\right ], 
\eary
and 
\bary
T_u&=&-\frac{g^2_W}{4M^2_W}
\left [
\left (1+ \frac{3m^2}{2M^2_W} +\frac {E^2_{\nu}}{M^2_W} \right )
(N_e-{\bar N}_e)\right.
\nonumber\\
&&\left. -\frac{2 E_{\nu}}{M^2_W}\Phi_3
-\frac{m^2 E_{\nu}}{M^2_W}\Phi_2
\right ].
\eary
In a heat bath one has to consider the thermal average of the
quantities. So we have to replace $E_{\nu}$ by 
$\langle E_{\nu}\rangle$ and $E_e$ by $\langle E_{e}\rangle$ in the medium. 
Using the above definitions, 
Eqs.(\ref{tkk}) and (\ref{tuu}) can be written as
\bary
T_k&=&-\frac{g^2_W}{4M^2_W} \left [
E_{\nu}\left (1+\frac{3}{2}\frac{m^2}{M^2_W}\right) 
(N_e-{\bar N}_e)\right.\nonumber\\
&&\left. -\frac{8E_{\nu}}
{3M^2_W} 
\left (\Ee N_e +\Eebar{\bar N}_e\right )
+\frac{2m^2{E_{\nu}}^2}{3M^2_W}
\left (\iEe N_e + \iEebar{\bar N}_e\right )
\right ],
\eary
and
\bary
T_u&=&-\frac{g^2_W}{4M^2_W} \left [
\left (
1+\frac{3}{2}\frac{m^2}{M^2_W}+\frac{E_{\nu}^2}{M^2_W} \right )
(N_e-{\bar N}_e)\right.\nonumber\\
&&\left. 
-\frac{2 E_{\nu}}{M^2_W}\left (\Ee N_e +\Eebar{\bar N}_e\right )
- \frac{m^2 E_{\nu}}{M^2_W} 
\left (\iEe N_e + \iEebar {\bar N}_e\right )
\right ].
\eary
Now for massless neutrino Eqs.(\ref{aw1}) and (\ref{bw1}) give,
\bary
a_W&=&-\sqrt{2} \frac{G_F}{M^2_W} 
\left [
E_{\nu}(N_e-{\bar N}_e)
+\frac{2}{3}\left (\Ee N_e +\Eebar{\bar N}_e\right ) \right.\nonumber\\
&&\left.
-\frac{5m^2}{3}
\left (\iEe {N_e} + \iEebar {\bar N}_e\right ) 
\right ],
\eary
and 
\bary
b_W&=&-\sqrt{2} G_F \left [ \left (1+\frac{3}{2} 
\frac{m^2}{M^2_W}\right )(N_e-{\bar N}_e) 
-\frac{8}{3} \frac{E_{\nu}}{M^2_W} 
\left (\Ee N_e +\Eebar{\bar N}_e\right )\right.\nonumber\\
&&\left.
+\frac{2}{3} \frac{m^2 E_{\nu}}{M^2_W} 
\left (\iEe {N_e}+ \iEebar {\bar N}_e\right )
\right ],
\eary
where $g^2_W/4M^2_W=\sqrt{2} G_F$, and $G_F$ is the Fermi coupling constant. 
Compared to $b_W$, $a_W$ is supressed by a factor $1/M^2_W$. 
Now we have to calculate the neutrino self-energy due to Z-boson exchange. 
The Z-coupling to the fermions is given by
\be
{\cal L}_Z=-i\frac{g}{2\cos\theta_W} {\bar f}
 \gamma^{\mu} \left ( C^f_V-C^f_A\gamma^5\right )f Z_{\mu},
\ee
where $C^f_V$ and $C^f_A$ denote the vector and axial vector couplings, which 
are given by
\be
C^f_V=\left \{\begin{array}
{r@{\quad\quad}l}
-\frac{1}{2}+2\sin^2\theta_W & e\\
\frac{1}{2} & {\nu}\\ \frac{1}{2}-2\sin^2\theta_W & {\text{p}}\\
-\frac{1}{2} & {\text{n}}
\end{array}\right.,
\ee
and 
\be
C^f_A=\left \{\begin{array}
{r@{\quad\quad}l}
\frac{1}{2} & {\nu},{\text{p}}\\
-\frac{1}{2} & e,{\text{n}}
\end{array}\right..
\ee
The contribution from the Z-exchange diagram Fig.\ref{fig1} (b) , 
can be obtained from 
the W-exchange diagram (a) by changing $M_W\rightarrow M_Z$.
The background here is made of neutrinos and anti-neutrinos 
instead of electrons and positrons. Making the above substitutions we obtain
\be
a_Z=-\frac{\sqrt{2} G_F}{M^2_Z} 
\left [
\Enu (N_{\nu}-{\bar N}_{\nu})
+\frac{2}{3}\Enu (N_{\nu}+{\bar N}_{\nu})
\right ],
\label{az}
\ee
and
\be
b_Z=-\sqrt{2} G_F \left [(N_{\nu}-{\bar N}_{\nu})
-\frac{8}{3} \frac{\Enu^2}{M^2_Z}(N_{\nu}+{\bar N}_{\nu}) 
\right ].
\label{bz}
\ee
Also $a_Z$ is suppressed by a factor $1/M^2_Z$ compared to $b_Z$.
Finally, we have to calculate the tadpole diagram due to Z-coupling  
Fig.\ref{fig1} (c). The contribution to the self-energy due to this 
diagram is given by
\bary
 -i\Sigma_{Z_t}(k)&=&
\frac{-ig}{2\cos\theta_W}\gmu
\left ( C^{\nu}_V - C^{\nu}_A\gamma^5\right )iD_{\mu\nu}(q=0)
\int \frac{d^4p}{(2\pi)^4} (-1)\times\nonumber\\
&\times&\text{Tr}
\left [
\frac{-ig}{2\cos\theta_W}\gnu\left ( C^{f}_V - C^{f}_A\gamma^5\right )iS_F(p)
\right ],
\eary
where $Z_t$ corresponds to the tadpole diagram due to Z-boson coupling.
The real part of this diagram is given by
\be
Re\Sigma_{Z_t}(k) =\frac{g^2}{M^2_Z\cos^2\theta_W} R \int \frac{d^3p}{(2\pi)^3}
\frac{1}{2E} C^f_V {\not p} (f(p.u) -{\bar f}(p.u)).
\ee
Finally this gives
\be
a_{Z_t}=0 ~~~~\text{and}~~~~ b_{Z_t}=-2\sqrt{2} G_F \sum_{f} \frac{C^f_V}{g} 
(N_f -{\bar N}_f).
\ee
This result implies that we have to consider the tadpole diagram contribution for different backgrounds obtaining
\bary
b_{Z_t}(e)&=&-\sqrt{2} G_F \left ( -\frac{1}{2} + 2 \sin^2\theta_W\right ) 
(N_e -{\bar N}_e),\nonumber\\
b_{Z_t}(\nu)&=&-\sqrt{2} G_F \sum_{\nu} (N_{\nu} -{\bar N}_{\nu}),\nonumber\\
b_{Z_t}(p)&=&
-\sqrt{2} G_F \left ( \frac{1}{2} - 2 \sin^2\theta_W\right ) 
(N_p -{\bar N}_p),\nonumber\\
b_{Z_t}(n)&=& 
\frac{\sqrt{2}}{2} G_F (N_n -{\bar N}_n),
\eary
for a background made of electrons, neutrinos, protons and 
neutrons respectively. For matter containing electrons, nucleons and 
neutrinos, the tadpole contribution will give
\bary
b_{Z_t} &=&-\sqrt{2} G_F 
\left [
\left ( -\frac{1}{2} + 2 \sin^2\theta_W\right ) (N_e -{\bar N}_e)
+\left ( \frac{1}{2} - 2 \sin^2\theta_W\right ) (N_p -{\bar N}_p)\right.
\nonumber\\
&&\left. -{\frac{1}2}(N_n -{\bar N}_n)+(N_{\nu_e} -{\bar N}_{\nu_e})+
(N_{\nu_{\mu}} -{\bar N}_{\nu_{\mu}})+(N_{\nu_{\tau}} -{\bar N}_{\nu_{\tau}})
\right ].
\eary
If the heat bath is charge neutral then,
\be
(N_p -{\bar N}_p)=(N_e -{\bar N}_e),
\ee
and this gives
\be
b_{Z_t}=\frac{G_F}{\sqrt{2}}\left [
(N_n -{\bar N}_n)
-2 \sum_{\nu}(N_{\nu} -{\bar N}_{\nu})\right ].
\ee
The test particle is moving in a heat bath and the particles (fermions) in 
the heat bath have average energy
\be
\langle E \rangle 
=\frac{g \int \frac{d^3p}{(2\pi)^3} f(E) E}{N_f},
\ee
where $N_f$ is the fermion number density already defined in Eq.(\ref{fand}).
In the non-relativistic limit ($\beta m \gg 1$), 
Eqs.(\ref{nd}) to (\ref{nddd}) reduces to
\bary
N_f&=&g \left (\frac{m}{2\pi\beta}\right )^{3/2} e^{-\beta (m-\mu)},\nonumber\\
{\cal J}_2&=& \frac{N_f}{\langle E\rangle} ~~~~\text{and} ~~~~
{\cal J}_3={\langle E\rangle} N_f.
\eary
But for the relativistic fermions $E\simeq |{\bf p}|$ and 
$N_f$ can be written as 
\be
N_f=2\int \frac{d^3p}{(2\pi)^3} f(E)=\frac{1}{\pi^2}\sum^{\infty}_{n=0}
(-1)^n \int^{\infty}_{0} E^2 dE e^{(n+1)\beta E}=\frac{3}{2\pi^2}
\xi(3) T^3,
\label{fnd1}
\ee
and similarly for photon, this is given by
\be
N_{\gamma}=2\int \frac{d^3p}{(2\pi)^3} f_{\gamma}(E)
=\frac{2}{\pi^2}\xi(3) T^3,
\ee
where $\xi(3)=1.20206$.
This gives, $N_e/N_{\gamma}=3/4$ and $N_{\nu}/N_{\gamma}=3/8$.
Then proceeding in the same way as in Eq.(\ref{fnd1}) we obtain
\be
2 \int \frac{d^3p}{(2\pi)^3} f(E) E=
\frac{1}{\pi^2}\sum^{\infty}_{n=0}
(-1)^n \int^{\infty}_{0} E^3 dE e^{(n+1)\beta E}=\frac{21}{4\pi^2}
\xi(4) T^4,
\ee
where $\xi(4)=\pi^4/90$ and we obtain the quantities
\be
\langle E_f\rangle = \frac{7}{2} \frac{\xi(4)}{\xi(3)} T ,
\ee
and 
\be 
\iEf\simeq \frac{1}{ \langle E_f\rangle}.
\ee
One can define the asymmetry parameter for the particle species $\alpha$ as
\be
L_{\alpha}\equiv\frac{N_{\alpha}-{\bar N}_{\alpha}}{N_{\gamma}}, 
\ee
where ${\alpha}$ can be lepton or baryon. Let us assume that the average 
particle and anti-particle energies are the same. Then using these we can write
\be
a_W=-\sqrt{2}\frac{G_F N_{\gamma}}{M^2_W}
\left [
\frac{7}{2} \frac{\xi(4)}{\xi(3)} T (1+ L_e) - \frac{5}{7} 
\frac{\xi(4)}{\xi(3)} \frac{m^2}{T}
\right ]
\ee
and
\be
b_W=-\sqrt{2}G_F N_{\gamma}
\left [
\left (1+\frac{3}{2}\frac{m^2}{M^2_W}\right )  L_e 
-\left (\frac{7\xi(4)}{\xi(3)}\right )^2 \left ( \frac{T}{M^2_W}\right )^2 +
\frac{m^2}{M^2_W}
\right ].
\ee
Similarly Eqs.(\ref{az}) and (\ref{bz}) can be written as
\be
a_Z=-\sqrt{2}\frac{G_F N_{\gamma}}{M^2_Z}
\left(\frac{7\xi(4)}{2\xi(3)}\right )
(1+L_{\nu_e}) T
\ee
and
\be
b_Z=-\sqrt{2} G_F N_{\gamma}
\left [
L_{\nu_e} -\frac{1}{2} 
\left(\frac{7\xi(4)}{\xi(3)}\right )^2 \frac{T^2}{M^2_Z}
\right ].
\ee
Finally the tadpole diagram contributions to the matter containing
$e, p, n$ and $\nu$ is
\be
b_{Z_t}= -\sqrt{2} G_F N_{\gamma} 
\left [
\left (-\frac{1}{2}+2 \sin^2\theta_W \right ) (L_e-L_p)
-\frac{L_n}{2}+L_{\nu_e}+L_{\nu_\mu}+L_{\nu_\tau}
\right ].
\ee
For the charge-neutral matter we have $ L_e=L_p $,
which gives
\be
b_{Z_t}=\sqrt{2}G_F N_{\gamma}
\left[
\frac{L_n}{2}-(L_{\nu_e} + L_{\nu_\mu} + L_{\nu_\tau})
\right ].
\ee
We have calculated the quantities $a$ and $b$ as given in Eq.(\ref{ls})
for all the three diagrams in Fig.\ref{fig1} and using these one can
calculate the neutrino effective potential, which is shown in the next section.
Also depending on the nature of the medium, there can be contributions from 
different backgrounds to $a$ and $b$. 

\section{Neutrino effective potential}
We have discussed earlier that, the Dirac equation for the neutrino 
is modified in the medium and this will give the dispersion relation 
\be
\text{det}\left ( (1+a){\not k} + b {\not u}\right )=0,
\ee
where the $a$ and $b$ are the same as given in Eq.(\ref{ls}) and det is 
the determinant. From this equation we obtain,
\be
(1+a) \omega + b = \pm (1+a) |{\bf k}|.
\ee
As we have seen earlier, $a$ is much smaller than $b$, so we can neglect the
$a$ contribution to the dispersion relation. Then
the dispersion relation for the particle (positive energy solution) is 
\be
(\omega -|{\bf k}|)\simeq -b.
\ee
From this we can see that, $-b $ is the effective potential
which will be experienced by the neutrino when propagating through a medium. 
So we have the effective potential
\be
V_{eff}\simeq -b.
\ee
For example, let us consider an electron neutrino propagating in a 
($i$) background of only electrons and positrons and ($ii$) only neutrinos
and anti-neutrinos.
In the first case the contribution to $b$ will be given by
\be
b=b_W + b_{Z_t}(e),
\ee
and in the second case
\be
b=b_Z +b_{Z_t}(\nu_e)+ b_{Z_t}(\nu_{\mu})+b_{Z_t}(\nu_{\tau}).
\ee

Let us consider two different physical situations
which can be realized in the cosmological and astrophysical scenarios, the
early universe hot plasma, before the BBN$^{3,4}$ and secondly the 
GRB fireball$^{13,14,15}$. Just before the 
nucleosynthesis when temperature was much above the electron mass but below the
muon mass i.e. $m \ll T \ll m_{\mu}$. In this situation, the contributions
will come from $e^{\pm}$, $n$, $p$ and all the neutrinos. Then the 
effective potential experience by a electron neutrino is given by
\bary
V_{\nu_e}&\simeq&\sqrt{2} G_F N_{\gamma}
\left [
\left (\frac{1}{2}+2 \sin^2\theta_W \right ) L_e
+\left (\frac{1}{2}-2 \sin^2\theta_W \right ) L_p \right.\nonumber\\
&&\left. -\frac{1}{2} L_n +2L_{\nu_e} 
+L_{\nu_\mu}+L_{\nu_\tau}-\left (1+\frac{1}{2}\cos^2\theta_W \right )
\left ( \frac{7\xi(4)}{\xi(3)}\right )^2 \frac{T^2}{M^2_W}
\right ].
\label{poteu}
\eary
If the test particle is ${\bar \nu}_e$, in the above Eq.(\ref{poteu}), 
the particle anti-particle asymmetry $L_i$ will be replaced by $-L_i$.
Secondly, let us consider the GRB fireball as a 
thermalized radiation and electron-positron plasma of temperature about
2 to 10 MeV. We also consider the situation that, the fireball is
contaminated by baryons$^{16,17,18}$. Thus ${\nu_e}$ propagating 
through it will experience an effective potential  
\be
V_{\nu_e,{\bar\nu_e}}=\sqrt{2} G_F N_{\gamma}\left [\pm {\cal L}
-\left ( \frac{7\xi(4)}{\xi(3)}\right )^2 \frac{T^2}{M^2_W}
\right ],
\ee
where
\be
{\cal L}=
\left (\frac{1}{2}+2 \sin^2\theta_W \right ) L_e + 
\left (\frac{1}{2}-2 \sin^2\theta_W \right ) L_p-\frac{1}{2}L_n.
\ee
If the fireball is charge neutral $i.e.$ electron asymmetry is the same as the
proton asymmetry ($L_e=L_p$) then 
\be
V_{\nu_e,{\bar\nu_e}} 
=\sqrt{2} G_F N_{\gamma}\left [\pm L_p \mp \frac{1}{2}L_n -
\left ( \frac{7\xi(4)}{\xi(3)}\right )^2 \frac{T^2}{M^2_W}
\right ].
\ee
The effective potential for $\nu_{\mu}$ and $\nu_{\tau}$ can also
 be calculated accordingly.

\section{Neutrino oscillation}
Here we consider the oscillation of neutrinos in the above mentioned two 
scenarios.
The evolution equation for propagation of neutrinos in the medium is given by
\be
i\left( 
\begin{array}{ccc} 
\dot{\nu_x}\\  
\dot{\nu_y}   
\end{array}
\right)= \left(
\begin{array}{ccc} 
V-\Delta \cos2\theta 
& \frac{\Delta}{2}\sin2\theta\\
\frac{\Delta}{2} \sin2\theta & 0  
\end{array}
\right)
\left(
\begin{array}{ccc} 
\nu_x\\ \nu_y
\end{array}
\right)
\label{mix} 
\ee
where $x$ is active and $y$ can be active or sterile, 
$\Delta=(m_2^2-m_1^2)/2E_{\nu}=\delta m^2/2E_{\nu}$, 
$V$ is the potential 
difference between $V_x$ and $V_y$, (i.e. $V=V_x-V_y$),
$E_{\nu}$ is the neutrino energy and $\theta$ is the mixing angle. 
The conversion probability for the above processes is given by,
\bary
{\cal P}_{x\rightarrow y}(t)&=&\frac{\Delta^2\sin^2 2\theta}
{(V-\Delta \cos 2\theta)^2+\Delta^2 \sin^2 2\theta} \times \nonumber\\
&\times&
\sin^2
\left ({\frac{\sqrt{(V-\Delta \cos 2\theta)^2
+\Delta^2 \sin^2 2\theta}}{2}}t\right ).
\label{prob}
\eary
In the above equation, the resonance condition is given by
\be
V=\frac{\delta m^2}{2E_{\nu}} \cos2\theta.
\label{reso}
\ee
So the resonance condition crucially depends on the sign of the potential.
In the early universe case, if we consider the 
$\nu_e\leftrightarrow\nu_{\mu}$ oscillation, then the effective potential is 
\be
V=V_{\nu_e}-V_{\nu_{\mu}}=
\sqrt{2} G_F N_{\gamma} \left [L_e+L_{\nu_e}-L_{\nu_{\mu}}
-\left (\frac{7\xi(4)}{\xi(3)}\right )^2 \frac{T^2}{M^2_W} \right ].
\label{poteup}
\ee
It is obvious from the above equation that the active-active oscillation is
independent of the baryonic contribution to the potential, whereas, if it is
active-sterile oscillation, then the potential does depend on the baryonic
component. Let us further assume that, $L_{\nu_e}=L_{\nu_{\mu}}$. This
assumption is probably reasonable, because, during this epoch all the neutrinos
are massless and equally populated. If this was the case, then 
$L_e > 6.14\times 10^{-9} T^2_{6}$ ($T_6$ is in units of $10^6$ eV) is needed 
to have resonant conversion along with the condition Eq.(\ref{reso}).  
As already discussed by
Enqvist et. al.,$^3$ the above oscillation can proceed equally from 
both sides and then
it will not create any asymmetry in electron and muon neutrinos. 
The anti-neutrinos oscillation will be very much
suppressed, due to the change in sign of the potential. But oscillation of 
active neutrinos to sterile ones will create asymmetry in $\nu$, ${\bar\nu}$ 
and this oscillation will depend on both lepton asymmetry and 
the baryon asymmetry in the medium.

The cosmological gamma ray bursts release about $10^{52}$ erg energy in a 
few second and  make them the most luminous object in 
the universe$^{13}$. The sudden 
release of a large amount of gamma-ray photons into a compact region, can lead 
to an opaque photon-lepton fireball through the production of electron-positron
pairs. It is believed that, the initial temperature of the fireball is in the
range 3 to 10 MeV, so pair creation takes place and it forms a lepton-photon
fireball with a typical radius of $10^7-10^8$ cm. The fireball is
contaminated by baryons both from the source as well as from the 
surrounding environment$^{16,17,18}$. Neutrinos produced within the fireball
due to inverse beta decay, or neutrinos from the progenitor, can 
propagate through the fireball. These propagating neutrinos can oscillate
resonantly or non-resonantly, depending on the nature of the fireball$^{19}$.  
For $\nu_e\leftrightarrow\nu_{\mu}$ oscillation, the potential is 
\be
V=V_{\nu_e}-V_{\nu_{\mu}}=
\sqrt{2} G_F N_{\gamma} \left [
L_e-\left (\frac{7\xi(4)}{\xi(3)}\right )^2 \frac{T^2}{M^2_W} \right ].
\label{potgrb}
\ee
If the asymmetries $L_{\nu_e}$= $L_{\nu_{\mu}}$ is valid in the early universe
plasma then it is exactly the same as in the  GRB-fireball. So we can say
that, GRB can mimic the early universe. 
One can see that the neutrino potential does not depend on the
baryon content of the fireball. The reason is, proton and neutron
contributions to the effective potential are the same for 
$\nu_e$ and $\nu_{\mu}$
and because of the difference it cancels out. So even if there are baryons in
the fireball, they will not affect the $\nu_e\leftrightarrow\nu_{\mu}$ or
the corresponding anti neutrino oscillation processes. Apart from this we
can also see that, for the antineutrino process, the potential is negative, 
hence the resonance condition Eq.(\ref{reso}) can never be satisfied. 
On the other hand
for the $\nu_e\leftrightarrow\nu_{\mu}$ process, depending on the value
of $L_e$, there can be a resonance. First of all, for resonance to occur,
the condition $L_e > 6.14\times 10^{-9} T^2_{6}$ has to satisfy and 
this is only possible for neutrino processes. Secondly one has to explore 
the neutrino energy, $\delta m^2$ and mixing angle of the neutrinos to
satisfy the resonance condition in Eq.(\ref{reso}).
  For example if we take $T_6=3$ then for resonance to take place 
$L_e >  5.6\times 10^{-8}$. Apart from this if we further assume a 
spherical fireball of radius $\sim$ 100 Km and $L_e=L_p$ then the 
fireball is contaminated by a mass $> 1.7\times 10^{-10} M_{\odot}$. 
In the fireball also,(as in the case of Early Universe) the oscillation 
of active neutrino to sterile one depends on the baryon and lepton contain 
of the fireball. A detail analysis of the neutrino oscillation in the GRB 
fireball is in progress.

\section{Conclusions}
We have calculated the $1/M^4$ order contribution to the neutrino 
self energy in Early Universe hot plasma and the GRB fireball, where the 
higher order contribution to the
neutrino effective potential has potential importance. We have looked for the 
condition of resonant conversion/oscillation of active to active neutrinos, 
which does not depend on the baryon content of the propagating medium. On the 
other hand, the active to sterile oscillation/conversion does depend on the 
 baryon content of the medium. It is shown that, under certain 
circumstances, the GRB fireball can behave as Early Universe hot plasma. The 
resonant oscillation of neutrino in the GRB fireball depends on its physical
parameters (temperature and particle asymmetry) as well as on the neutrino
mixing angle and mass square differences. If this conditions are satisfied, 
the resonant oscillations of active to active neutrinos may have an effect 
on the energy release of the GRB Fireball.  This has to be studied 
in great detail in future.

This work is partially supported by the grant from PAPIIT-UNAM IN119405.\\

\end{document}